# Deep Learning based Direct Segmentation Assisted by Deformable Image Registration for Cone-Beam CT based Auto-Segmentation for Adaptive Radiotherapy

Xiao Liang, Howard Morgan, Ti Bai, Michael Dohopolski, Dan Nguyen, and Steve Jiang[*]

Medical Artificial Intelligence and Automation Laboratory and Department of Radiation Oncology, University of Texas Southwestern Medical Center, Dallas, TX, USA

*Corresponding author.

Corresponding author address:
Steve Jiang
Department of Radiation Oncology
UT Southwestern Medical Center, Dallas, TX
2280 Inwood Road, Dallas, TX 75390
Phone: 214-648-8510
Email: steve.jiang@utsouthwestern.edu

**Abstract**

Cone-beam CT (CBCT)-based online adaptive radiotherapy calls for accurate auto-segmentation to reduce the time cost for physicians to edit contours, since the patient is immobilized on the treatment table waiting for treatment to start. However, deep learning (DL)-based direct segmentation of CBCT images is a challenging task, mainly due to the poor image quality and lack of well-labelled large training datasets. Deformable image registration (DIR) is often used to propagate the manual contours on the planning CT (pCT) of the same patient to CBCT. In this work, we undertake solving the problems mentioned above with the assistance of DIR. Our method consists of three main components. First, we use deformed pCT contours derived from multiple DIR methods between pCT and CBCT as pseudo labels for initial training of the DL-based direct segmentation model. Second, we use deformed pCT contours from another DIR algorithm as influencer volumes to define the region of interest for DL-based direct segmentation. Third, the initially trained DL model is further fine-tuned using a smaller set of true labels. We found that DL-based direct segmentation on CBCT trained with pseudo labels and without influencer volumes shows poor performance compared to DIR-based segmentation. However, adding deformed pCT contours as influencer volumes in the direct segmentation network dramatically improves segmentation performance, reaching the accuracy level of DIR-based segmentation. The DL model with influencer volumes can be further improved through fine-tuning using a smaller set of true labels. Experiments showed that 7 out of 19 structures have an at least 0.2 Dice similarity coefficient increase compared to DIR-based segmentation. A DL-based direct CBCT segmentation model can be improved to outperform DIR-based segmentation models by using deformed pCT contours as pseudo labels and influencer volumes for initial training, and by using a smaller set of true labels for model fine tuning.

# 1. Introduction

Online adaptive radiotherapy (ART) is an advanced radiotherapy technology in which the daily treatment plan is adapted to the patient's changing anatomy (e.g., shrinking tumor, losing body weight), typically based on cone beam computed tomography (CBCT) images. The online nature of the treatment demands high efficiency since the patient is immobilized while waiting for treatment to start. The time-consuming process of segmenting the tumor volumes and organs at risk (OARs) has become a major bottleneck for the widespread use of online ART, warranting an urgent need for accurate auto-segmentation tools (Glide-Hurst et al., 2021).

Auto-segmentation of CBCT images is a very challenging task, mainly due to poor image quality and lack of training labels for deep learning (DL)-based methods. First, the greater presence of noise and artifacts on CBCT images, such as capping, cupping, ring, and streaking artifacts, makes CBCT more difficult than CT for auto-segmentation tasks (Lechuga and Weidlich, 2016). Second, contouring of tumor volumes and OARs is not part of the common applications of CBCT-based image-guided radiotherapy. Therefore, unlike CT for treatment planning, one cannot use the clinical contours generated from routine clinical practice to train DL models for CBCT segmentation. Expert clinicians must retrospectively contour large sets of CBCT images specifically for CBCT segmentation research, which is time consuming and challenging. Due to two major limitations, poor image quality and lack of a well-labeled large set of training data, studies have shown that DL-based direct segmentation on CBCT images produce poor segmentation results (Alam et al., 2021; Beekman et al.; Dahiya et al., 2021; Dai et al., 2021; Léger et al., 2020).

Auto-segmentation on CBCT for ART is a unique task when planning CT (pCT) with manual contours are available. Using pCT with manual contours as prior knowledge, some studies have shown that DL-based direct segmentation could achieve improved results. A simple way to take advantage of pCT and its contours is to directly mix them into a limited CBCT training dataset. This cross-domain augmentation of the training set was effective for CBCT augmentation (Léger et al., 2020). With a more complicated data augmentation strategy, one study generated variant synthetic CBCT (sCBCT) images with one pair of pCT and CBCT of the same patient where the generated data was used to train a CBCT segmentation model (Dahiya et al., 2021). Similarly, some other studies utilized artifact induction to convert pCT to sCBCT to make use of high quality CT manual contours for CBCT segmentation (Alam et al., 2021; Schreier et al., 2020).

All the studies mentioned above use data augmentation methods to mitigate the lack of training labels on CBCT, either by adding pCT with manual contours into the training set directly or by generating sCBCT with contour labels from pCT with manual contours. While auto-segmentation results can be improved to some degree by data augmentation, the biggest drawback of these methods is that neither CT nor sCBCT can truly represent a real CBCT image. Therefore, a more robust and popular way to utilize pCT and its high quality contours for CBCT auto-segmentation is through deformable image registration (DIR) methods. By deforming pCT to CBCT, pCT contours can be propagated to CBCT. While traditional DIR methods including B-spline and Demons algorithms (Fedorov et al., 2012; Gu et al., 2009; Klein et al., 2010) are computationally intensive and time consuming, DL-based DIR methods can be quick at inference, but usually require large amounts of training data (Han et al., 2021), which can be mitigated with methods like test-time optimization (TTO) (Liang et al., 2022). The biggest benefit of DIR-based segmentation is that the topological consistency of contours can be preserved by smoothly deforming pCT contours, since most organs have small anatomical changes. However, when anatomical changes are big, DIR-based contour propagation can be biased towards the original pCT contour shape.

A combination of DL-based direct segmentation and DIR-based contour propagation can potentially leverage the advantages of two methods. One way is to firstly use a DL-based model to direct segment easier OARs in CBCT images, then subsequently use the segmentation results to constrain the DIR between pCT and CBCT, and finally to propagate the target volumes and rest of OARs from pCT to CBCT (Archambault et al., 2020). Another way is through joint learning of segmentation and registration. A typical way for joint segmentation and registration is to predict segmentations on both moving and fixed images using unsegmented moving and fix images as inputs (Estienne et al., 2019; Xu and Niethammer, 2019). Another way for joint learning is to use the moving and fixed image, as well as moving segmentations as inputs to predict segmentations on the fixed image (Beekman et al., 2021). In those approaches for joint learning, DL-based direct segmentation and DIR-based contour propagation are combined either through parameter sharing between the segmentation and registration model, or through a joint loss function. However, the above-mentioned joint learning approaches cannot significantly outperform DIR-based contour propagation for CBCT segmentation, and still requires a large amount of segmentation labels on fix (CBCT) images for model training (Beekman et al., 2021).

In this paper, we explore a new method to improve DL-based direct segmentation with the assistance of DIR results, aiming to outperform the DIR-based methods for CBCT segmentation, without requiring a large well-labeled training dataset. We propose to use pseudo labels for initial model training, where the pseudo labels are deformed pCT contours. To help localize OARs and target, we propose to add deformed pCT contours as influencer volumes through additional channels of the segmentation model. We then fine-tune the initially trained model using a small set of training data with true labels.

## 2. Methods

### 2.1. Problem definition

In a fully supervised segmentation task, we can denote the training set as $\mathcal{D} = \{(X, Y)\}^D$, where $X \in \mathbb{R}^\Omega$ denotes training images and $Y \in \{0,1\}^\Omega$ denotes their corresponding pixel-wise labels. $\Omega$ denotes its corresponding spatial domain. Given the labeled dataset $D$, the segmentation task intends to learn a function F with parameter $\theta$ to map $X$ to $Y$ by minimizing the standard Dice loss:

$$\min_\theta \mathcal{L}_{Dice}(\theta) := 1 - \frac{[2\sum_{p\in\Omega} y(p)s_\theta(p)]+c}{\sum_{p\in\Omega} y(p)+\sum_{p\in\Omega} s_\theta(p)+c}, \qquad (1)$$

where $s_\theta = F(x|\theta)$ is the predicted probabilities by the CNNs, and $c$ is a small constant added to prevent dividing by 0. $s_\theta \in [0,1]^\Omega$, with 0 and 1 denoting background and foreground.

### 2.2. Pseudo label learning

We generated pseudo labels for initial training of the model and firstly deformed pCT to its paired CBCT to get a deformation vector field (DVF). We then used DVF to warp pCT's contours to generate deformed contours, which were the pseudo labels of CBCT. To address the pseudo label noise, we applied multiple DIR algorithms to generate multiple sets of pseudo labels. By randomly selecting one type of pseudo label during each training iteration, we could mitigate random errors coming from DIR algorithms. We applied three different DL-based DIR algorithms: FAIM (Kuang and Schmah, 2019), 5-cascaded Voxelmorph (Dalca et al., 2019), and 10-cascaded VTN (Zhao et al., 2019) to generate pseudo labels $y_1$, $y_2$, and $y_3$, respectively. Given the dataset $\mathcal{D} = \{(X, Y)\}^D$, where $X = \{x\}$ represents image and $Y = \{y_i\}$ represents pseudo labels, the segmentation model intends to learn a function F with parameter $\theta_{pl}$ formulated similarly to Equation 1:

$$\min_{\theta_{pl}} \mathcal{L}_{Dice}(\theta_{pl}) := 1 - \sum_i \frac{\left[2\sum_{p\in\Omega} y_i(p) s_{\theta_{pl}}(p)\right] + c}{\sum_{p\in\Omega} y_i(p) + \sum_{p\in\Omega} s_{\theta_{pl}}(p) + c}, \tag{2}$$

where $s_{\theta_{pl}} = F(x|\theta_{pl})$ is the model prediction and $i \in \{1,2,3\}$.

### 2.3. Influencer volumes

To deal with the low image quality problem for direct CBCT segmentation, we proposed to add influencer volumes as additional channels of input, as shown in Figure 1. The architecture used in this experiment was a typical U-Net architecture. Besides the CBCT image, deformed pCT contours were used as additional input channels to constrain the region of interest for segmentation. The influencer volumes were used for shape and location feature extraction. The shape and location features were independently extracted from the influencer volumes for each level and combined with features extracted from CBCT images by multiplication. The combined features were further concatenated with features from up-sampling layers. The output consisted of multi-organ segmentation masks derived from learned relations between the CBCT images and influencer volumes.

Since deformed pCT contours were also used as pseudo labels for training, to avoid using the same deformed pCT contours as input and output at the same time, we proposed to assign two different deformed pCT contours as such by randomly picking two DIR methods to generate different deformed contours during each training iteration. In this case, the input was $X = \{x, y_j\}$ and label was $Y = \{y_i\}$, where $i, j \in \{1,2,3\}$ and $i \neq j$. We could formulate the loss function in a similar style to Equation 1 or Equation 2:

$$\min_{\theta_{iv}} \mathcal{L}_{Dice}(\theta_{iv}) := 1 - \sum_i \frac{\left[2\sum_{p\in\Omega} y_i(p) s_{\theta_{iv}}(p)\right] + c}{\sum_{p\in\Omega} y_i(p) + \sum_{p\in\Omega} s_{\theta_{iv}}(p) + c}, \tag{3}$$

where $s_{\theta_{iv}} = F(x, y_j | \theta_{iv})$ is the model prediction with image x and deformed pCT segmentation $y_j$ as input.

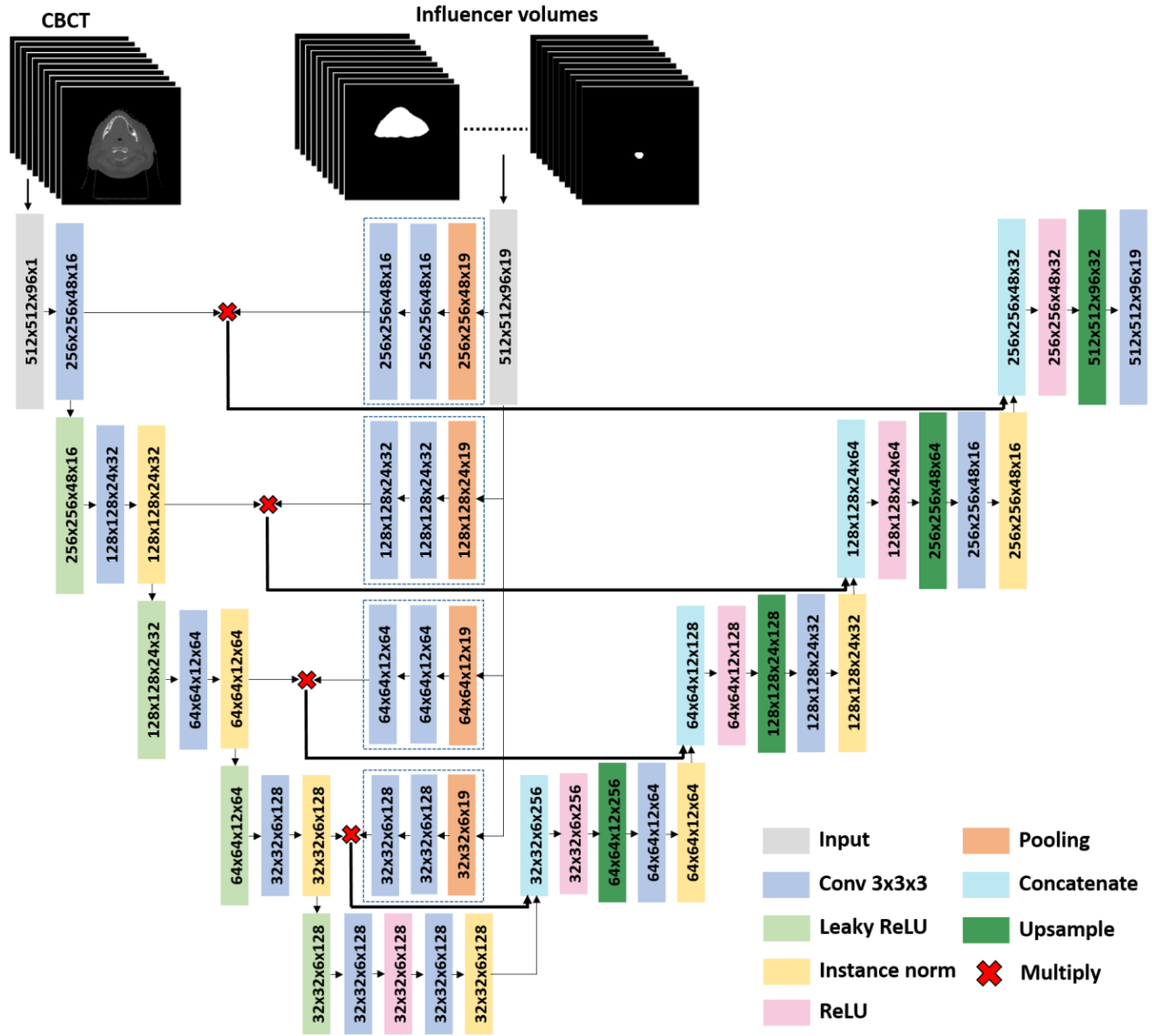

Figure 1. A U-Net architecture with deformed pCT contours as influencer volumes for CBCT segmentation.

### 2.4. Fine-tuning with true labels

To further improve the performance of the model initially trained with pseudo labels, we proposed to fine-tune the model using a small training dataset with true labels.

### 2.5. Data

We retrospectively collected data from 137 patients with head and neck (H&N) squamous cell carcinoma treated with conventionally fractionated external beam radiotherapy. Each patient's data included a 3D pCT volume acquired before the treatment course, OARs and target segmentations delineated and approved by radiation oncologists on the pCT, and a 3D CBCT image. The pCT volumes were acquired by a Philips CT scanner (Philips Healthcare, Best, Netherlands) with 1.17 × 1.17 × 3.00 mm$^3$ voxel spacing. The

CBCT volumes were acquired by Varian On-Board Imagers (Varian Medical Systems Inc., Palo Alto, CA, USA) with 0.51 × 0.51 × 1.99 mm$^3$ voxel spacing and 512 × 512 × 93 dimensions. Among those 137 patients, 39 patients had true segmentation labels on CBCT drawn by a radiation oncology expert. Nineteen structures that were either critical OARs or had large anatomical changes during radiotherapy courses were selected as segmentation targets. These structures were: left brachial plexus (L_BP), right brachial plexus (R_BP), brainstem, oral cavity, constrictor, esophagus, nodal gross tumor volume (nGTV), larynx, mandible, left masseter (L_Masseter), right masseter (R_Masseter), posterior arytenoid-cricoid space (PACS), left parotid gland (L_PG), right parotid gland (R_PG), left superficial parotid gland (L_Sup_PG), right superficial parotid gland (R_Sup_PG), left submandibular gland (L_SMG), right submandibular gland (R_SMG), and spinal cord.

To generate pseudo labels and influencer volumes, image registration was performed between pCT and CBCT for the 98 patients without true CBCT segmentation labels. pCT is first rigid registered to its corresponding CBCT through Velocity (Varian Medical Systems Inc., Palo Alto, CA, USA), and then deformedly registered to the CBCT through our previously proposed DIR methods (Liang et al., 2022): applying TTO to three different state-of-the-art DIR models including FAIM, 5-cascaded Voxelmorph, and 10-cascaded VTN. Subsequently, the pCT contours were warped accordingly to generate deformed pCT contours as pseudo labels or influencer volumes for training. The remaining 39 patients with true CBCT labels were grouped into 30 for model fine-tuning and nine for model testing. CBCT images and contour masks were padded to size of 512×512×96 and 512×512×96×19 from 512×512×93 and 512×512×93×19, respectively.

## 2.6. Experiments

Three experiments were performed for this work. First, we trained the U-Net model on the 98 patients with pseudo labels by switching the training label among three types of pseudo labels without any prior knowledge ($Model_{pseudo}$) or adding any influencer volumes to study the performance of the direct segmentation. Then, we added influencer volumes into U-Net and trained the network with pseudo labels to observe the performance gained from adding influencer volumes ($Model_{influencer}$). Both the influencer volumes and pseudo labels were deformed pCT contours, but coming from different DIR algorithms. Finally, we fine-tuned $Model_{BB}$ on 30 patients with true labels to further improve segmentation accuracy ($Model_{finetune}$). During the fine-tuning stage, we applied early stopping, layer freezing, and lower learning rate to prevent model overfitting. All three models were tested on nine patients and dice similarity coefficient (DSC) was used to evaluate segmentation accuracy. In this work, DIR-based contour propagating was used as the baseline, since it is the most commonly used method in current clinical practice for auto-segmentation in CBCT-based ART. We considered 10-cascaded VTN model with TTO as the state-of-the-art DIR baseline model ($Model_{DIR}$).

## 3. Results

### 3.1. Model trained on pseudo labels without influencer volumes

$Model_{pseudo}$ exhibited worse performance than $Model_{DIR}$, as shown in Table 1, with all 19 structures achieving lower DSC scores with $Model_{pseudo}$. This suggests that models depending on CBCT images alone cannot derive reliable segmentation results. The main reasons for these inferior outcomes are listed below based on our observations.

Firstly and most importantly, CBCT images have many artifacts and low soft tissue contrast compared to CT images. Like the images shown in Figure 2(a), some organs including the brainstem, esophagus, parotid gland, and submandibular gland do not have a clear boundary from surrounding tissues, making segmentation more difficult. However, those organs usually do not have significant anatomical changes, and the deformed pCT contours are quite accurate.

Secondly, for some OARs, superior and inferior ends sometimes produce large segmentation errors due to the direct segmentation model's inability to deal with certain geometry information if not provided with guidelines, as shown in Figure 2(b). For example, a consensus guideline for CT-based delineation of the constrictor (Brouwer et al., 2015) specified that the cranial border was defined as the caudal tip of pterygoid plates, and the caudal border as the lower edge of the cricoid cartilage. However, $Model_{pseudo}$ failed to pick up this border data from the training data directly, leading to delineation errors around the superior and inferior borders.

Thirdly, target volumes and some OARs are extremely challenging to segment, even on CT. Target delineation, like nodal clinical tumor volume, is more variable and therefore more difficult to predict than OARs. The brachial plexus is difficult to localize on CT images and is identified with adjacent structures using additional help from anatomic texts or magnetic resonance imaging. It is unsurprising to see that DIR-based segmentation prediction has fairly good performance, since it uses pCT contours as a start point. However, direct segmentation of those extremely difficult structures is prone to failure.

Fourthly, direct segmentation models are prone to inaccurate or incomplete labels in the training dataset. For example, in Figure 2(d), the manual contours are actually the left superficial parotid gland and spinal canal, but mislabeled as parotid gland and spinal cord, separately. In addition, some structures may be modified to represent avoidance structures and not necessarily hold fast to the exact anatomic boundaries of that organ, such as with the oral cavity where the portion overlapping the planning tumor volume is sometimes cropped out, leaving the manual segmentation of oral cavity incomplete. We also found that some organs lack complete delineation in the superior-inferior direction in the training dataset, because a complete delineation was unnecessary if a part of the organ was distant from the target volume.

Fifthly, outliers in the testing dataset have a negative impact on model performance. Figure 2(e) shows two outliers that we observed in the testing dataset. One test patient had a tracheostomy tube, however, no such patients are in the training dataset. The presence of the tracheostomy tube leads to incorrect delineation of the esophagus by the direct segmentation model. Another test patient had the esophagus pushed away to his left side, but no similar patient exists in the training dataset. $Model_{pseudo}$ usually has poor performance on outliers, while $Model_{DIR}$ is more accurate by preserving shape information from pCT contours.

(a)

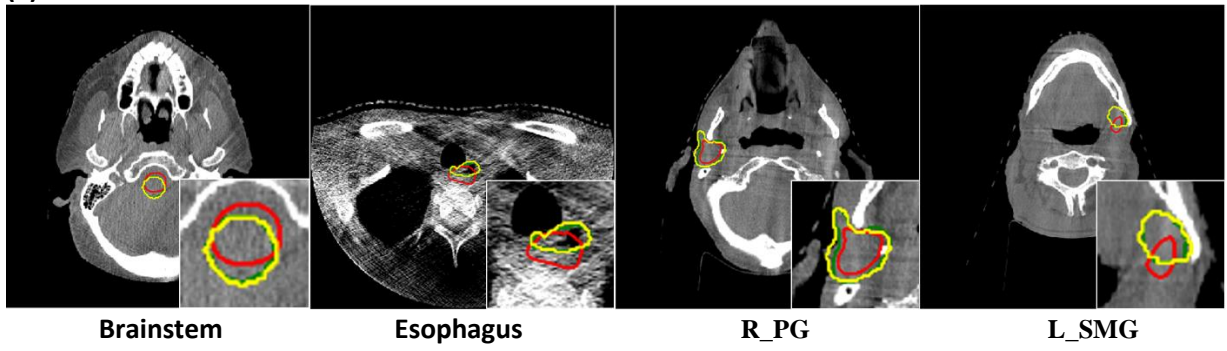

| Brainstem | Esophagus | R_PG | L_SMG |

(b)

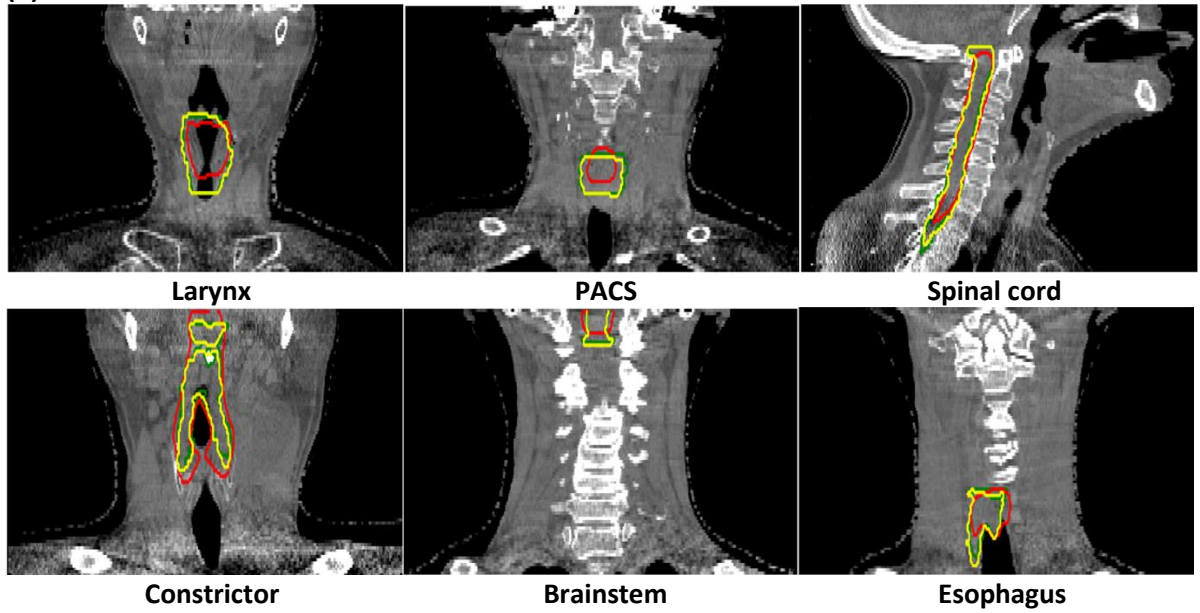

| Larynx | PACS | Spinal cord |
| Constrictor | Brainstem | Esophagus |

(c)

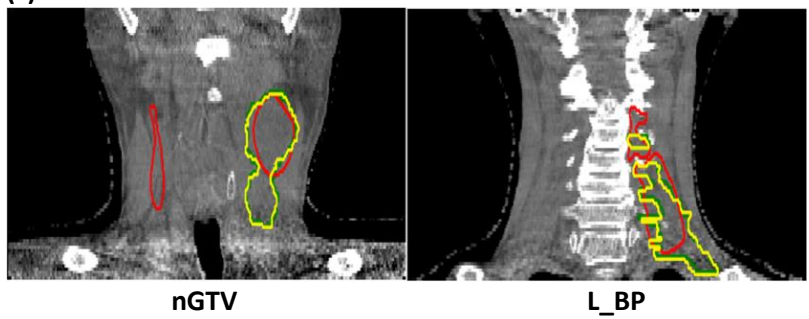

| nGTV | L_BP |

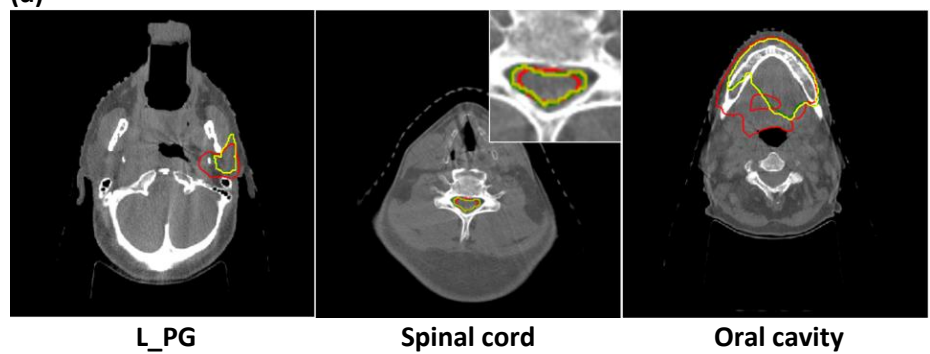

| L_PG | Spinal cord | Oral cavity |

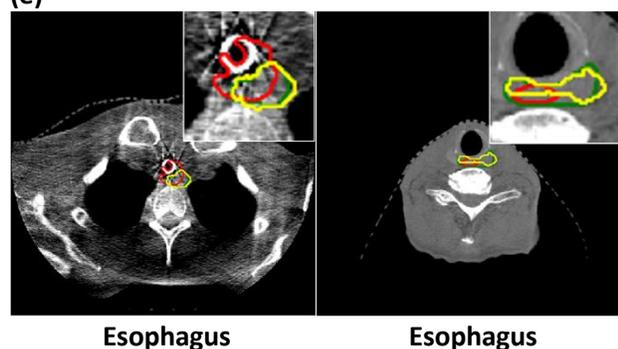

| Esophagus | Esophagus |

**Figure 2. Categories of reasons that cause poor performance of direct segmentation without any prior knowledge on CBCT.** Green lines are manual contours drawn by a radiation oncology expert, yellow lines are DIR propagated contours ($Model_{DIR}$), and red lines come from direct segmentation without influencer volumes ($Model_{pseudo}$).

### 3.2. Model trained on pseudo labels with influencer volumes

Table 1 shows that $Model_{DIR}$ and $Model_{influencer}$ have similar DSC scores over all 19 structures. With the use of influencer volumes from pseudo labels, the performance of the DS model can be significantly improved to the level of DIR-based segmentation. It is not surprising that the performance of $Model_{influencer}$ does not surpass that of $Model_{DIR}$, since $Model_{influencer}$ used the pseudo labels generated by DIR for training.

### 3.3 Model fine-tuned on real labels with influencer volumes

When $Model_{influencer}$ is fine-tuned using true labels, the DIR errors contained in pseudo labels for training could potentially be corrected, allowing for the model performance to surpass that of $Model_{DIR}$. Table 1 shows that DSC scores of $Model_{finetune}$ are greater than or equal to those of $Model_{DIR}$ and $Model_{influencer}$. $Model_{finetune}$ outperforms $Model_{DIR}$ for 7 out of 19 structures for at least a 0.02 DSC improvement, with the DSC improvement of a single structure up to 0.03. The average DSC over 19 structures by $Model_{finetune}$ is 0.86, with a minimum DSC of 0.72 for L_BP and maximum DSC of 0.95 for the oral cavity. Examples of segmentation from axial, frontal, and sagittal views by $Model_{DIR}$, $Model_{influencer}$, and $Model_{finetune}$ are shown in Figure 3 for visual evaluation. We can see that

$Model_{finetune}$ not only can maintain shape characteristics of the prior segmentation in pCT, it can also eliminate the errors caused by the prior segmentation.

**Table 1. Average DSC of 9 test patients for different auto-segmentation models.** $Model_{DIR}$ is DIR only segmentation. $Model_{pseudo}$ is direct DL segmentation using pseudo labels for training. $Model_{influencer}$ is direct DL segmentation using both pseudo labels and influencer volumes for training. $Model_{finetune}$ is derived from fine-tuning $Model_{influencer}$ with true labels. Numbers in green indicate that the DSC has at least a 0.02 improvement compared to $Model_{DIR}$, while numbers in red means the opposite.

| Structure | $Model_{DIR}$ | $Model_{pseudo}$ | $Model_{influencer}$ | $Model_{finetune}$ |
|---|---|---|---|---|
| L_BP | 0.71 | 0.34 | 0.71 | 0.72 |
| R_BP | 0.73 | 0.40 | 0.73 | 0.75 |
| Brainstem | 0.91 | 0.72 | 0.90 | 0.91 |
| Oral cavity | 0.95 | 0.60 | 0.95 | 0.95 |
| Constrictor | 0.83 | 0.72 | 0.85 | 0.86 |
| Esophagus | 0.83 | 0.62 | 0.83 | 0.83 |
| nGTV | 0.83 | 0.32 | 0.84 | 0.84 |
| Larynx | 0.89 | 0.73 | 0.89 | 0.90 |
| Mandible | 0.89 | 0.87 | 0.90 | 0.91 |
| L_Masseter | 0.90 | 0.84 | 0.90 | 0.92 |
| R_Masseter | 0.90 | 0.83 | 0.90 | 0.90 |
| PACS | 0.78 | 0.66 | 0.79 | 0.81 |
| L_PG | 0.89 | 0.70 | 0.88 | 0.89 |
| R_PG | 0.91 | 0.72 | 0.90 | 0.91 |
| L_Sup_PG | 0.84 | 0.60 | 0.84 | 0.84 |
| R_Sup_PG | 0.86 | 0.50 | 0.85 | 0.86 |
| L_SMG | 0.84 | 0.59 | 0.85 | 0.86 |
| R_SMG | 0.84 | 0.55 | 0.85 | 0.85 |
| Spinal cord | 0.86 | 0.75 | 0.86 | 0.89 |

**(a) R_BP**

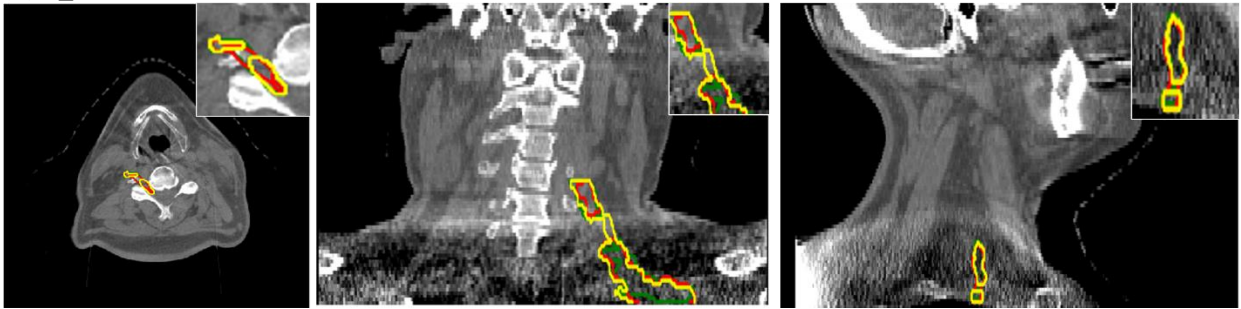

**(b) Constrictor**

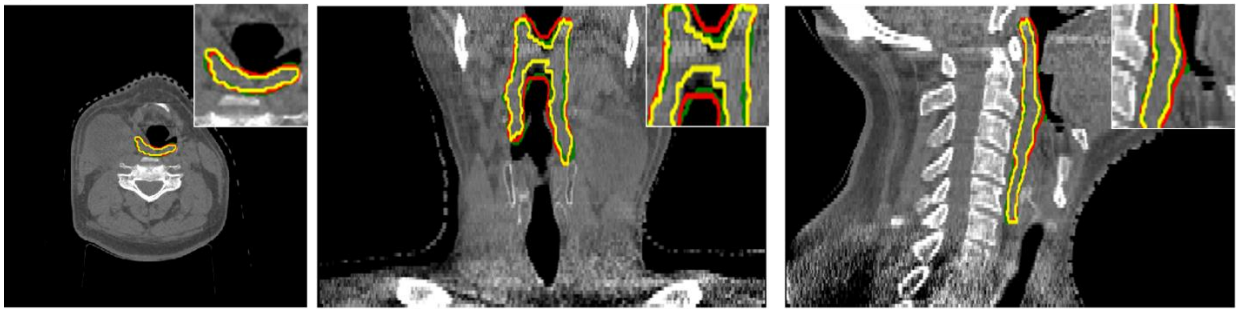

**(c) Mandible**

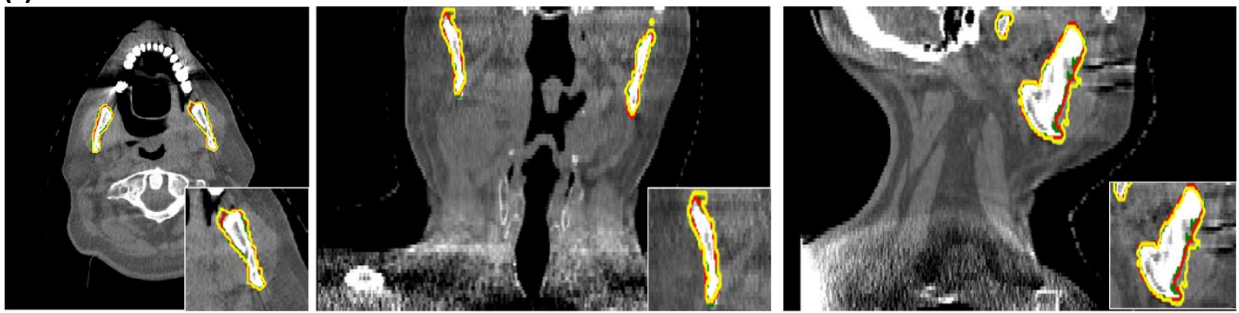

**(d) L_Masseter**

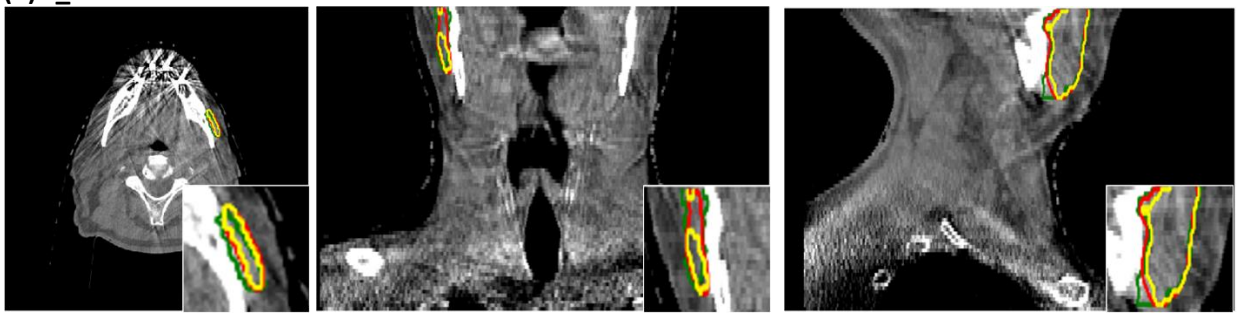

**(e) PACS**

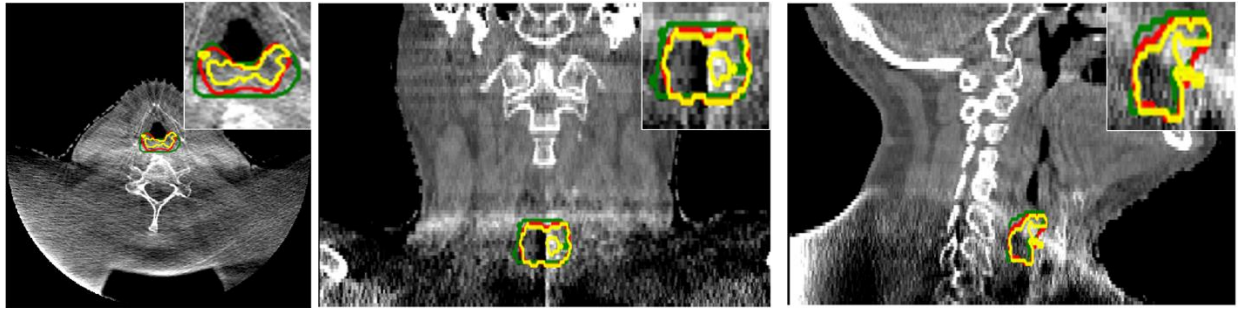

**(f) L_SMG**

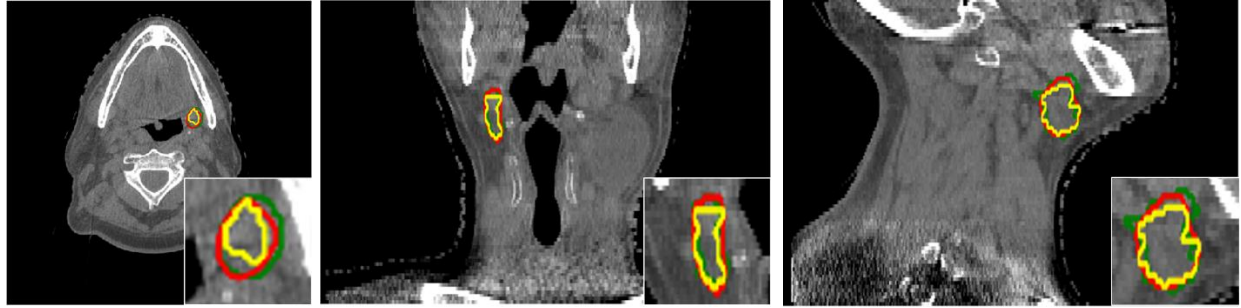

**(g) Spinal cord**

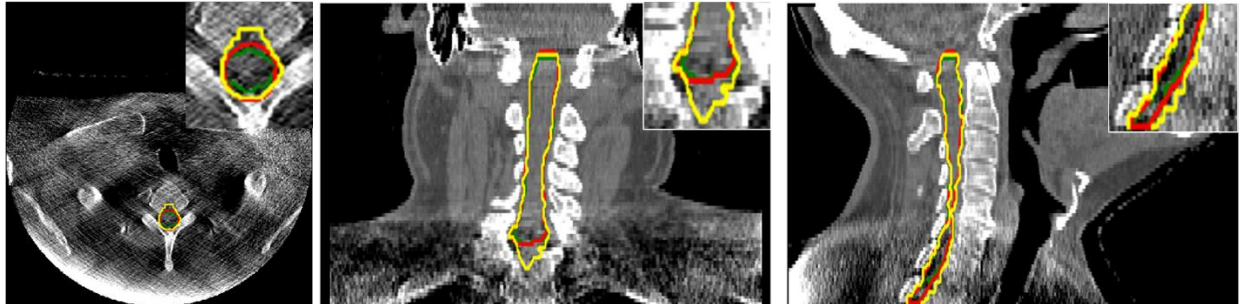

**Figure 3. Segmentation examples in axial, frontal, and sagittal views.** Green lines are manual contours drawn by a radiation oncology expert, yellow lines are DIR-based segmentation ($Model_{DIR}$), and red lines are segmentations from fine-tuned direct segmentation with influencer volumes ($Model_{finetune}$).

## 4. Discussion and Conclusions

Based on this work, it is evident that direct segmentation on CBCT images without prior knowledge is infeasible, mainly due to the poor image quality, superior and inferior border uncertainty, delineation complexity, outliers, inaccurate or incomplete labels, and also lack of true labels. With pCT and its corresponding contours available in the ART workflow as prior knowledge, the accuracy of DL-based direct CBCT segmentation can be greatly improved.

Different from CT auto-segmentation tasks where large amount of labels are usually available for training, manual labels are not common on CBCT images. To solve this lack of a large well-labelled training dataset, we proposed to use deformed pCT as pseudo labels for the initial DL model training, as the performance of the initially trained model is far inferior to the DIR-based methods. We then proposed to use deformed pCT from another DIR algorithm as influencer volumes in the network. By adding influencer volumes as new channels to the model to constrain shape and localization, the model performance can be dramatically improved, reaching the level of DIR-based methods. To outperform the DIR-based methods, the DL-based direct segmentation model initially trained with pseudo labels and influencer volumes can be fine-tuned using a small set of training data with true labels.

Fine-tuning with true labels could mitigate DIR errors contained in the pseudo labels since in the fine-tuning stage, there are several ways to prevent overfitting (Ying, 2019). Reducing overfitting by training the network on more datasets is not considered in this work since we have only have a small amount of data with true labels. Reducing overfitting by changing the complexity of the network is another way to prevent overfitting. For example, the model could be tuned via freezing some layers and only updating parameters of the remaining layers. Another simple alternative to avoid overfitting is to improve regularization (Goodfellow et al., 2016) by early stopping via monitoring model performance on a validation set and stopping training when performance degrades. Meanwhile, adding regularization requires a smaller learning rate.

We used H&N patients to test our models, since CBCT-based ART is often used for this disease site and since segmentation is more challenging. The same approach can be easily expanded to and tested on datasets from other disease sites.

In summary, to overcome the two major issues related to CBCT based image segmentation for online ART, such as poor image quality and lack of well-labelled large training datasets, we developed a method to use DIR-propagated contours as pseudo labels and influencer volumes for initial training and subsequently fine-tuned the model using a small set of a training dataset with true labels. The method has been tested with a cohort of H&N cancer patients and demonstrated superior segmentation accuracy to the commonly used DIR-based methods.

## Acknowledgement

We would like to thank the Varian Medical Systems Inc. for supporting this study and Ms. Sepeadeh Radpour for editing the manuscript.

## Conflict of Interest Statement

The authors declare no competing financial interest. The authors confirm that all funding sources supporting the work and all institutions or people who contributed to the work, but who do not meet the criteria for authorship, are acknowledged. The authors also confirm that all commercial affiliations, stock ownership, equity interests or patent licensing arrangements that could be considered to pose a financial conflict of interest in connection with the work have been disclosed.

## Reference

Alam, S.R., Li, T., Zhang, P., Zhang, S.-Y., Nadeem, S., 2021. Generalizable cone beam CT esophagus segmentation using physics-based data augmentation. Physics in Medicine & Biology 66, 065008.
Archambault, Y., Boylan, C., Bullock, D., Morgas, T., Peltola, J., Ruokokoski, E., Genghi, A., Haas, B., Suhonen, P., Thompson, S., 2020. Making on-line adaptive radiotherapy possible using artificial intelligence and machine learning for efficient daily re-planning. Med Phys Intl J 8.

Beekman, C., van Beek, S., Stam, J., Sonke, J.-J., Remeijer, P., Improving predictive CTV segmentation on CT and CBCT for cervical cancer by diffeomorphic registration of a prior. Medical Physics n/a.
Beekman, C., van Beek, S., Stam, J., Sonke, J.-J., Remeijer, P., 2021. Improving predictive CTV segmentation on CT and CBCT for cervical cancer by diffeomorphic registration of a prior. Medical Physics n/a.
Bishop, C.M., 1995. Neural networks for pattern recognition. Oxford university press.
Brouwer, C.L., Steenbakkers, R.J.H.M., Bourhis, J., Budach, W., Grau, C., Grégoire, V., van Herk, M., Lee, A., Maingon, P., Nutting, C., O'Sullivan, B., Porceddu, S.V., Rosenthal, D.I., Sijtsema, N.M., Langendijk, J.A., 2015. CT-based delineation of organs at risk in the head and neck region: DAHANCA, EORTC, GORTEC, HKNPCSG, NCIC CTG, NCRI, NRG Oncology and TROG consensus guidelines. Radiotherapy and Oncology 117, 83-90.
Dahiya, N., Alam, S.R., Zhang, P., Zhang, S.-Y., Li, T., Yezzi, A., Nadeem, S., 2021. Multitask 3D CBCT-to-CT translation and organs-at-risk segmentation using physics-based data augmentation. Medical Physics 48, 5130-5141.
Dai, X., Lei, Y., Wang, T., Dhabaan, A.H., McDonald, M., Beitler, J.J., Curran, W.J., Zhou, J., Liu, T., Yang, X., 2021. Head-and-neck organs-at-risk auto-delineation using dual pyramid networks for CBCT-guided adaptive radiotherapy. Physics in Medicine & Biology 66, 045021.
Dalca, A.V., Balakrishnan, G., Guttag, J., Sabuncu, M.R., 2019. Unsupervised learning of probabilistic diffeomorphic registration for images and surfaces. Medical Image Analysis 57, 226-236.
Estienne, T., Vakalopoulou, M., Christodoulidis, S., Battistella, E., Lerousseau, M., Carre, A., Klausner, G., Sun, R., Robert, C., Mougiakakou, S., Paragios, N., Deutsch, E., 2019. U-ReSNet: Ultimate Coupling of Registration and Segmentation with Deep Nets, MICCAI 2019: Medical Image Computing and Computer Assisted Intervention – MICCAI 2019, Shenzhen, China, pp. 310-319.
Fedorov, A., Beichel, R., Kalpathy-Cramer, J., Finet, J., Fillion-Robin, J.-C., Pujol, S., Bauer, C., Jennings, D., Fennessy, F., Sonka, M., Buatti, J., Aylward, S., Miller, J.V., Pieper, S., Kikinis, R., 2012. 3D Slicer as an image computing platform for the Quantitative Imaging Network. Magnetic Resonance Imaging 30, 1323-1341.
Glide-Hurst, C.K., Lee, P., Yock, A.D., Olsen, J.R., Cao, M., Siddiqui, F., Parker, W., Doemer, A., Rong, Y., Kishan, A.U., Benedict, S.H., Li, X.A., Erickson, B.A., Sohn, J.W., Xiao, Y., Wuthrick, E., 2021. Adaptive Radiation Therapy (ART) Strategies and Technical Considerations: A State of the ART Review From NRG Oncology. International Journal of Radiation Oncology*Biology*Physics 109, 1054-1075.
Goodfellow, I., Bengio, Y., Courville, A., 2016. Deep learning. MIT press.
Gu, X., Pan, H., Liang, Y., Castillo, R., Yang, D., Choi, D., Castillo, E., Majumdar, A., Guerrero, T., Jiang, S.B., 2009. Implementation and evaluation of various demons deformable image registration algorithms on a GPU. Physics in Medicine and Biology 55, 207-219.
Han, X., Hong, J., Reyngold, M., Crane, C., Cuaron, J., Hajj, C., Mann, J., Zinovoy, M., Greer, H., Yorke, E., Mageras, G., Niethammer, M., 2021. Deep-learning-based image registration and automatic segmentation of organs-at-risk in cone-beam CT scans from high-dose radiation treatment of pancreatic cancer. Medical Physics 48, 3084-3095.
Klein, S., Staring, M., Murphy, K., Viergever, M.A., Pluim, J.P.W., 2010. <emphasis emphasistype="mono">elastix</emphasis>: A Toolbox for Intensity-Based Medical Image Registration. IEEE Transactions on Medical Imaging 29, 196-205.
Kuang, D., Schmah, T., 2019. Faim–a convnet method for unsupervised 3d medical image registration, International Workshop on Machine Learning in Medical Imaging. Springer, pp. 646-654.
Lechuga, L., Weidlich, G.A., 2016. Cone Beam CT vs. Fan Beam CT: A Comparison of Image Quality and Dose Delivered Between Two Differing CT Imaging Modalities. Cureus 8, e778-e778.

Léger, J., Brion, E., Desbordes, P., De Vleeschouwer, C., Lee, J.A., Macq, B., 2020. Cross-Domain Data Augmentation for Deep-Learning-Based Male Pelvic Organ Segmentation in Cone Beam CT. Applied Sciences 10.

Liang, X., Chun, J., Morgan, H., Bai, T., Nguyen, D., Park, J.C., Jiang, S., 2022. Segmentation by Test-Time Optimization (TTO) for CBCT-based Adaptive Radiation Therapy. arXiv preprint arXiv:2202.03978.

Schreier, J., Genghi, A., Laaksonen, H., Morgas, T., Haas, B., 2020. Clinical evaluation of a full-image deep segmentation algorithm for the male pelvis on cone-beam CT and CT. Radiotherapy and Oncology 145, 1-6.

Xu, Z., Niethammer, M., 2019. DeepAtlas: Joint Semi-supervised Learning of Image Registration and Segmentation, In: Shen, D., Liu, T., Peters, T.M., Staib, L.H., Essert, C., Zhou, S., Yap, P.-T., Khan, A. (Eds.), Medical Image Computing and Computer Assisted Intervention – MICCAI 2019. Springer International Publishing, Cham, pp. 420-429.

Ying, X., 2019. An Overview of Overfitting and its Solutions. Journal of Physics: Conference Series 1168, 022022.

Zhao, S., Dong, Y., Chang, E.I., Xu, Y., 2019. Recursive cascaded networks for unsupervised medical image registration, Proceedings of the IEEE/CVF International Conference on Computer Vision, pp. 10600-10610.